\documentclass[10pt,conference]{IEEEtran}

\IEEEoverridecommandlockouts
\usepackage{adjustbox}
\usepackage{longtable}
\usepackage{cite}
\usepackage{underscore}
\usepackage{amsmath,amssymb,amsfonts}
\usepackage{algorithmic}
\usepackage{graphicx}
\usepackage{textcomp}
\usepackage{xcolor}
\usepackage[euler]{textgreek}
\usepackage{color}
\usepackage{url}
\usepackage{multirow}
\usepackage{tabularx}
\usepackage{subfig}
\usepackage{enumitem}
\usepackage{caption}
\usepackage{float}
\usepackage{pdfcomment}
\usepackage[utf8x]{inputenc}

\DeclareUnicodeCharacter{2212}{-}

\begin{document}

\renewcommand\ttdefault{cmvtt}

\title{\texttt{CONTINUER}: Maintaining Distributed DNN Services During Edge Failures}

\author{\IEEEauthorblockN{Ayesha Abdul Majeed\IEEEauthorrefmark{1}, Peter Kilpatrick\IEEEauthorrefmark{1}, Ivor Spence\IEEEauthorrefmark{1}, and Blesson Varghese\IEEEauthorrefmark{1}\IEEEauthorrefmark{2}}
\IEEEauthorblockA{\IEEEauthorrefmark{1}\textit{School of Electronics, Electrical Engineering and Computer Science, 
Queen's University Belfast, UK}\\
E-mail: \{aabdulmajeed01, p.kilpatrick, i.spence\}@qub.ac.uk}
\IEEEauthorblockA{\IEEEauthorrefmark{2}\textit{School of Computer Science, 
University of St Andrews, UK}\\
E-mail: bv6@st-andrews.ac.uk}
}

\maketitle
\thispagestyle{plain}
\pagestyle{plain}

\begin{abstract}

Partitioning and deploying Deep Neural Networks (DNNs) across edge nodes may be used to meet performance objectives of applications. However, the failure of a single node may result in cascading failures that will adversely impact the delivery of the service and will result in failure to meet specific objectives. The impact of these failures needs to be minimised at runtime. Three techniques are explored in this paper, namely repartitioning, early-exit and skip-connection. When an edge node fails, the repartitioning technique will repartition and redeploy the DNN thus avoiding the failed nodes. The early-exit technique makes provision for a request to exit (early) before the failed node. The skip connection technique dynamically routes the request by skipping the failed nodes. This paper will leverage trade-offs in accuracy, end-to-end latency and downtime for selecting the best technique given user-defined objectives (accuracy, latency and downtime thresholds) when an edge node fails. To this end, \texttt{CONTINUER} is developed. Two key activities of the framework are estimating the accuracy and latency when using the techniques for distributed DNNs and selecting the best technique. It is demonstrated on a lab-based experimental testbed that \texttt{CONTINUER} estimates accuracy and latency when using the techniques with no more than an average error of 0.28\% and 13.06\%, respectively and selects the suitable technique with a low overhead of no more than 16.82 milliseconds and an accuracy of up to 99.86\%.  
\end{abstract}


\section{Introduction}
\label{sec:introduction}

Distributed Deep Neural Network (DNN) services that underpin many modern applications are known to benefit from edge computing~\cite{lockhart2020scission,kang2017neurosurgeon}. The responsiveness of applications can be improved because edge computing offers compute resources to process data closer to the source where it is generated. Pretrained DNN models are adopted by real-time image recognition applications such as virtual reality gaming or autonomous vehicles, where devices continuously transmit the data to the cloud for inference. To achieve distributed DNN inference, DNNs are partitioned by the granularity of neural network layers. As a result, DNN layers that demand a substantial amount of processing power can be offloaded to the cloud or to edge servers. 


The failure of one resource, such as a compute node in an edge computing environment, can result in cascading failures. This adversely impacts the DNN service and may violate application requirements, such as latency thresholds~\cite{deng2021intelligent}. Traditionally, techniques, such as checkpointing, container migration, and server replication have been adopted to reduce the impact of failures~\cite{tang2018migration,containerMigration}. However, they are not best suited for edge environments in which resources may be compute/storage limited and are highly decentralised. 

One technique to minimise the impact of failures specific to distributed DNN services is repartitioning~\cite{majeed2021neukonfig}. 
A monolithic DNN can be partitioned and its layers are distributed across multiple resources to meet performance objectives of the application~\cite{wang2019adda}. 
If a node fails, then the DNN can be repartitioned such that a new distribution of the DNN layers is generated so that the layers originally mapped on to the failed edge node can be executed elsewhere. 
While DNN repartitioning will maintain the original accuracy, it is expensive in that it has both computational and communication overheads making it unsuitable as a general technique for mitigating failures in any edge environment. 

Alternate techniques, namely early-exit~\cite{branchynet} and skip connection~\cite{resnet32} that exploits certain characteristics of the DNN to maintain the service of a distributed DNN when an edge node fails are explored in this paper. 
While these techniques have been adopted in the literature to meet runtime requirements of applications (considered further in Section~\ref{sec:relatedwork}), but have been minimally considered in the context of edge failures. 
In the early-exit technique, a classification request can be terminated early before it reaches a failed edge node. While early-exit reduces the inference latency, it impacts the accuracy of the DNN.
In the skip connection technique, the route of a classification request can be dynamically adapted to skip failed nodes. However, adopting skip connection reduces the latency but requires extra resources and increases bandwidth use.

This paper develops a novel framework, namely \texttt{CONTINUER} that maintains the service of distributed DNNs when edge outages occur. Failure recovery strategies for edge services have been relatively less studied compared to failure detection~\cite{2019infrastructure,aral2018dependency}. Therefore, \texttt{CONTINUER} focuses on service recovery from failures by presenting a mechanism to recover from a service failure. \texttt{CONTINUER} selects one of the three techniques (repartitioning, early-exit and skip connection) by accounting for: (i) trade-offs in accuracy and performance characteristics (such as latency and downtime\footnote{Downtime in this paper refers to the time taken to recover from edge failure.} incurred) for each technique, and (ii) user-defined objectives, such as accuracy, latency and downtime thresholds. 

\texttt{CONTINUER} operates in two phases. The first is the profiler phase in which the accuracy and latency of a DNN can be estimated for the three techniques with low overhead to respond quickly to failures. The second is the runtime phase in which a scheduler selects the best technique for an edge failure given the estimated accuracy and latency of a DNN for the three techniques and user-defined objectives. 
Using a lab-based testbed comprising two processor platforms and two production DNNs, namely ResNet-32 and MobileNetV2 it is demonstrated that \texttt{CONTINUER} (i) estimates the accuracy and latency of DNNs for the three techniques with an average percentage error of no more than 0.28\% and 13.06\%, respectively, and (ii) selects a suitable technique when an edge node fails with a low overhead of no more than 16.82~milliseconds and high accuracy of up to 99.86\%. 

The remainder of this paper is organised as follows: Section~\ref{sec:background} presents the techniques used for maintaining the service of a DNN when there is an edge failure.
Section~\ref{sec:framework} presents the \texttt{CONTINUER} framework, including the design considerations and the operating phases. Section~\ref{sec:profiler} presents both the profiler and runtime phases of \texttt{CONTINUER}.
Section~\ref{sec:experiments} presents the experimental results obtained from a lab-based testbed to confirm the feasibility of \texttt{CONTINUER}.
Section~\ref{sec:relatedwork} presents the related work. 
Section~\ref{sec:conclusions} concludes the paper by considering the limitations of the current work and future directions of this research.

\section{Background}
\label{sec:background}

This section presents the context of our work together with two DNNs that are used as case studies.

\subsection{Edge Failures}
Service outages at the edge can occur as a result of intermittent network connectivity, and link or node failure~\cite{aral2020learning}. To ensure the continued availability of services, mechanisms such as check-pointing~\cite{containerMigration}, replication~\cite{tang2018migration} and rescheduling of services~\cite{aral2018dependency} have been used. Since a distributed DNN is deployed over different edge nodes to provide collaborative inference~\cite{distributedexit}, the failure of a single node may have a cascading effect and cause the failure of other nodes. 

When an edge node fails, DNN applications deployed on the edge will need to be redeployed. Redeploying the failed services by repartitioning the DNN incurs a downtime making it unsuitable for latency-critical applications. Alternative approaches to ensure the resilience of a DNN service in the face of edge outages exploit certain characteristics of the DNN. For example, dynamically adapting to routes within the DNN such that failed nodes are skipped (referred to as skip-connection) or terminating requests before reaching a failed node. 

\subsection{Underpinning Techniques}
The techniques used in \texttt{CONTINUER} to ensure continued availability of services are repartitioning, early-exit and skip connections. While the repartitioning technique has been employed in the literature to adapt a distributed DNN
for performance efficiency (layers of the DNN are distributed across multiple resources, such as the edge and the cloud), it has not been employed in the context of edge node failures. Similarly, early-exit and skip connection are two techniques that are employed in the literature for reducing delays in inference to effectively use limited compute resources that may be available~\cite{matsubara2021split}. For the first time, these techniques will be utilised to ensure that a distributed DNN can carry on providing services when an edge outage occurs.

\subsubsection{Technique 1 -- Repartitioning}
A DNN model comprising a sequence of layers can be partitioned and distributed over the edge and the cloud to meet privacy and performance objectives (such as end-to-end inference latency) 
 of an application. Inference requests generated from a device may be processed by the first partition (the initial sequence of layers of the DNN) on the edge. The intermediate results are transferred via the network to the second partition (the remaining sequence of layers of the DNN) on the cloud. 
The performance of the partition executing on the edge may vary due to the system load (such as CPU or memory utilization) and the overall performance can be affected due to an unstable network between the edge and the cloud. The system must adapt to these changes that occur at runtime to maintain the required performance of the DNN application. This is achieved by repartitioning (finding a different layer at which the DNN should be partitioned) and redeploying the partitions on the edge and the cloud~\cite{lockhart2020scission, mcnamee2021case}. In the repartitioning technique, another edge-cloud pipeline can be employed to deploy the distributed DNN to reduce the service downtime on the edge~\cite{majeed2021neukonfig}. 


\subsubsection{Technique 2 -- Early-exit} The second technique is early-exit, which as the name implies makes provision for an inference request to exit (early) before the last layer of the DNN. This allows for accelerating inference albeit there may be accuracy losses. To this end, the base DNN architecture is required to be modified and intermediate classifiers are added to the layers after which inference requests can exit. The classifiers added to the base model allow an input sample to be classified in the intermediate layers~\cite{branchynet, distributedexit}.

Production DNNs comprising many layers will be large and have substantial computational requirements. This makes it challenging to deploy such DNNs on the edge where there may be resource limitations, and if deployed the end-to-end latency will increase. Dynamic DNNs that make use of early-exit can reduce the end-to-end latency and can be adapted to suit the computational resources available during inference and the input characteristics provided to the DNN~\cite{matsubara2021split}. 

\subsubsection{Technique 3 -- Skip connection}
This technique facilitates the skipping of one or more layers of the DNN model.

A skip connection is achieved within a DNN by using gating networks that are inserted between layers in the DNN model. The gating networks map the output of a previous layer to a binary decision whether to enable or skip the next layer~\cite{skipnet}. Skip connections were designed to accelerate the training speed and improve the accuracy of large DNNs~\cite{resnet32}. Skip connections have been used to provide input-aware dynamic inference and reduce the computational cost of using large DNNs~\cite{wu2018blockdrop}.

\subsection{DNN Selection}
Two DNNs, namely ResNet-32 and MobileNetV2 are selected for investigation in the \texttt{CONTINUER} framework.

\subsubsection*{ResNet-32} The Residual Network (ResNet)~\cite{resnet32} is designed with skip connections to speed up the training process and to achieve a high accuracy. The ResNet model consists of residual blocks, made up of two or more convolutional layers, and skip-connections, which allow direct paths between any two residual blocks. A residual block is defined as
\begin{equation}
    y = F(x, \left \{Wi\right \}) + x
\label{eqn:resnet-block}
\end{equation}
where $x$ denotes the input and $y$ denotes the output vector, $F(x, {Wi})$ is the function for the residual mapping to be learned and $F + x$ is the operation performed by a shortcut connection and element-wise addition.

Residual blocks have been found not to have a strong relationship with each other~\cite{veit2016residual} and it is also noted that the classification errors increase when more residual blocks are skipped from the model during inference.

\textit{MobileNetV2:}
This is a DNN model  for compact, low-latency, low-power mobile devices\cite{sandler2018mobilenetV2}. The architecture consists of 17 residual blocks followed by a $1\times1$ convolution, a global average pooling layer, and a classification layer.

\subsection{Suitability of Underpinning Techniques for Edge Failures}

ResNet and MobileNetV2 are selected as the base DNN network because of their architectural design, which includes pre-defined skip connections. Assume a DNN is distributed over edge nodes $N=\left \{ n_{1},n_{2},n_{3},n_{4},n_{5} \right \}$ in a network. If a service outage occurs at $n_{3}$, then the path to $n_{4}$ and $n_{5}$ is disconnected, thereby preventing any further inference requests from being processed.

To ensure resilience
when nodes fail ($n_{3}$ in the example), the DNN can be redeployed over the first two nodes ($n_{1}$ and $n_{2}$). For this the first technique presented above, namely repartitioning can be employed. When the DNN is repartitioned, a new partitioning layer at which the DNN can be partitioned and deployed on $n_{1}$ and $n_{2}$ will be identified. This technique will achieve the same accuracy as the original DNN, but may incur a downtime for repartitioning~\cite{majeed2021neukonfig}.

Alternatively, the failed node ($n_{3}$) may either be bypassed by using the skip connection technique or the inference request can be terminated at $n_{2}$ by using the early-exit technique. Skip connection will address the edge failure problem by bypassing the node that failed. The early-exit technique addresses the edge failure problem by terminating incoming requests before the failed node. The skip connection technique may have a higher accuracy than early-exit, however, the relative end-to-end latency will be higher.

\section{The \texttt{CONTINUER} Framework}
\label{sec:framework}
In this section, the design of \texttt{CONTINUER} is  presented. The design considerations are: it should be agnostic to the infrastructure; decision making should be rapid with low overhead; and user-defined objectives should be accounted for.

\subsection{Assumption}
A DNN is represented as a directed acyclic graph with a set of layers $L = \{l_{1}, l_{2}, l_{3},...\}$ and a group of layers is termed as a block. The set of blocks is denoted by $B = \{b_{1}, b_{2}, b_{3},...\}$. An edge computing system is assumed that consists of a set of nodes $N=\left \{ n_{1},n_{2},n_{3},...\right \}$ over which a DNN model is distributed. It is also assumed that each block is placed on a different node as shown in Figure~\ref{fig:early-technique}.

\subsection{Technique Selection}
\begin{figure}[t]
    \centering
    \includegraphics[width=0.45\textwidth]{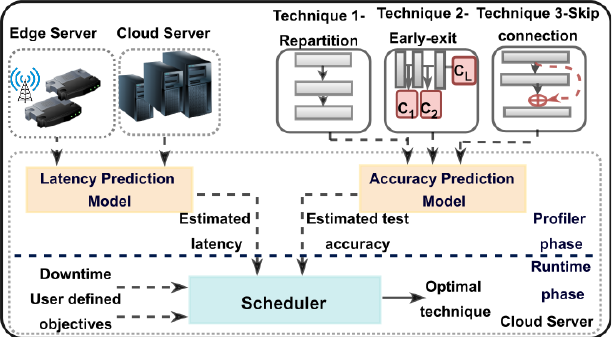}
    \caption{Overview of the \texttt{CONTINUER} framework}
    \label{fig:framework}
\vspace{-4.5mm}
\end{figure}

Figure \ref{fig:framework} illustrates the proposed \texttt{CONTINUER} framework which operates in two phases. 

\subsubsection*{Profiler phase} 
The profiler phase collects the values for the metrics on which \texttt{CONTINUER} relies and operates in offline mode. These metrics are the accuracy and the end-to-end latency of the DNN model and the downtime incurred when selecting a suitable technique when a node fails. There are two components within the profiler phase, namely the Accuracy Prediction Model and the Latency Prediction Model. In the profiler phase, the accuracy and latency of the DNN model are profiled for training the accuracy and latency prediction models. 
The Accuracy Prediction Model estimates the accuracy of the DNN at runtime if any of the three techniques (partitioning, early-exit and skip connection) were to be selected given the pretrained weights of the DNN can be provided as input.

The purpose of the Latency Prediction Model is to estimate the layer latency of the techniques given the layer hyperparameters of the DNN. It is not feasible to measure accuracy and end-to-end latency  at runtime using a profiling technique due to timing constraints and the speed required in selecting a technique to mitigate the impact of node failure and maintain the service of a DNN. Therefore, prediction models are employed that estimate accuracy and end-to-end latency during runtime. The profiler phase is discussed in Section~\ref{sec:profiler}.

\subsubsection*{Runtime phase} 
During the runtime phase the Scheduler determines the suitable technique that needs to be used given a node failure. The Scheduler takes as input the estimated accuracy, estimated end-to-end latency, and downtime (empirical) and selects the optimal technique for mitigating the impact of the node failure. During runtime, the value for the downtime metric is calculated as the time taken to predict and retrieve the estimated accuracy and end-to-end latency parameters.
The runtime phase is further discussed in Section~\ref{sec:profiler}.

\section{The Profiler and Runtime Phases}
\label{sec:profiler}
To determine a suitable technique to maintain the service of a distributed DNN when a node failure occurs, the \texttt{CONTINUER} framework makes use of three metrics, namely accuracy, latency, and downtime associated with each technique. 

\subsection{Profiler Phase} 
Resource-independent and resource-specific values are gathered in the profiler phase. The latency metric is a resource-specific value that depends on the underlying hardware. Accuracy and downtime are resource-independent, depending only on the DNN model and the technique (repartitioning, early-exit and skip connection), respectively. 
The values of accuracy and latency of the techniques are estimated in the profiler phase, which is carried out offline. Downtime values are however not estimated, instead empirical values of downtime of each technique are used.

In this section, first the partition points defined within the repartitioning technique and the modification of the DNN models for defining early exit points and skip connections are presented. Then the data collection approach and the prediction models used for estimating the latency and accuracy of each technique for different exit points and skip connections that can be used when a node fails are considered. Finally, the empirical values obtained for downtime metric are discussed.
 
\subsubsection{Repartitioning}
\begin{figure*}[t]
\begin{center}
	\subfloat[Partition points defined for ResNet-32]
	{\label{fig:ResNet-repartition}
	\includegraphics[width=0.75\textwidth]
	{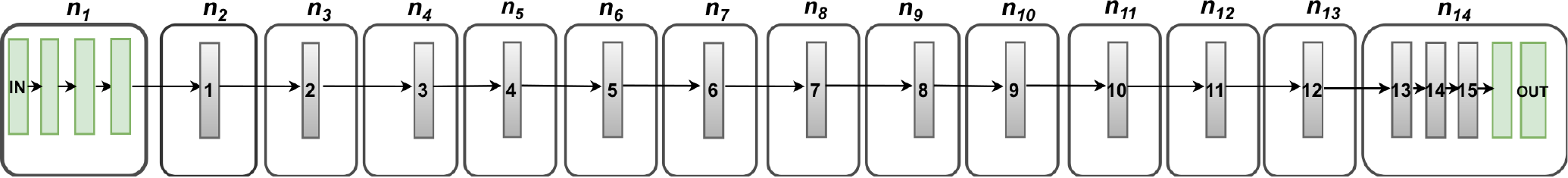}}
	
	\subfloat[Partition points defined for MobileNetV2]
	{\label{fig:MobileNet-repartition}
	\includegraphics[width=0.75\textwidth]
	{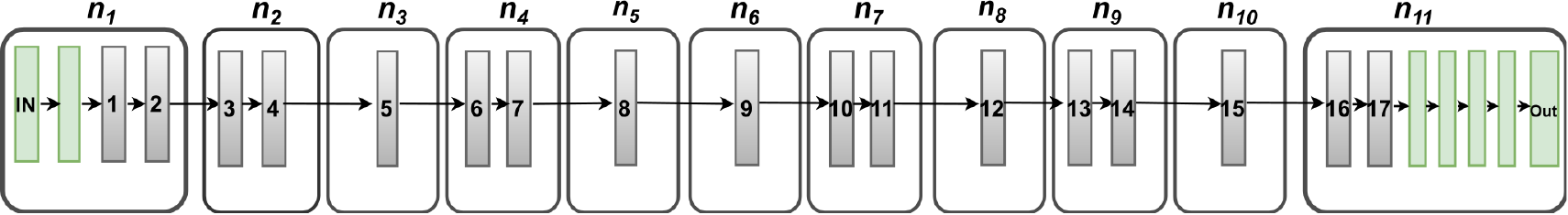}}
\end{center}
\caption{Distribution of partition points of ResNet-32 and MobileNetV2 when layers and block of layers are distributed across different compute nodes.}
\label{fig:repartition-node}
\vspace{-4.5mm}
\end{figure*}

In the repartitioning technique, the DNN is partitioned after each residual block represented by grey bars in Figure~\ref{fig:repartition-node}. The partitioned points are defined under the assumption that each residual block of DNN is placed on different nodes. The architecture of ResNet-32 consists of an
initial convolutional layer, batch normalisation and activation layers, 
15 residual blocks, followed by a global average pool layer and dense layer. The architecture of MobileNetV2 consists of 17 residual blocks, followed by a 1×1 convolution, a global average pooling layer, and a dense layer. ResNet-32 can be distributed on up to fourteen nodes (Figure~\ref{fig:ResNet-repartition}) and  MobileNetV2 on up to eleven nodes (Figure~\ref{fig:MobileNet-repartition}).

\subsubsection*{Model training parameters for Repartitioning}
For  repartitioning, the ResNet-32 model is trained with a learning rate of $1e-4$ and MobileNetV2 is trained with a learning rate of $1e-3$. Both models are trained using loss function categorical cross-entropy on the CIFAR-10 dataset\footnote{https://www.cs.toronto.edu/~kriz/learning-features-2009-TR.pdf} that contains 50000 training and 10000 test images of resolution 32×32 consisting of 10 classes, with a batch size of 64 and epoch size of 500. The epoch size is set to 500 to generate a dataset of 500 instances for the accuracy prediction model for predicting accuracy through pretrained weights. The accuracy obtained for ResNet-32 is 82.52\% and for MobileNetV2  85.54\%.

\subsubsection{Early-exit} 
\begin{figure*}[t]
\begin{center}
	\subfloat[Early exit points ($E_{1}-E_{13}$) defined for ResNet-32]
	{\label{fig:ResNet-early}
	\includegraphics[width=0.70\textwidth]
	{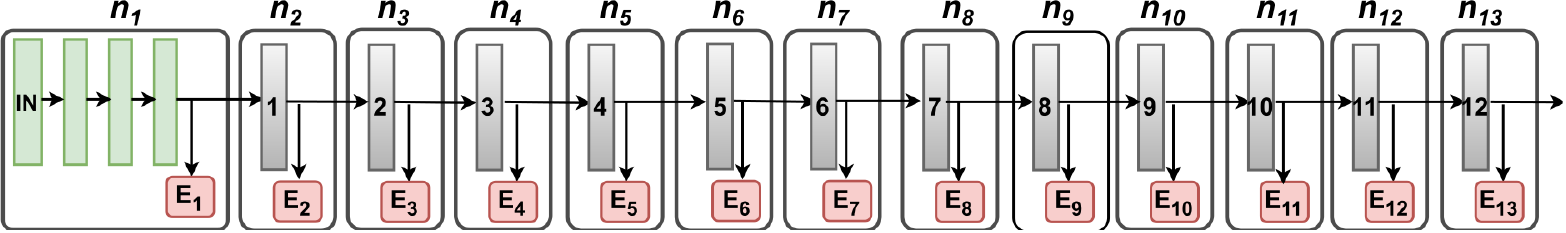}}
	
	\subfloat[Early exit points ($E_{1}-E_{10}$) defined for MobileNetV2]
	{\label{fig:MobileNet-early}
	\includegraphics[width=0.70\textwidth]
	{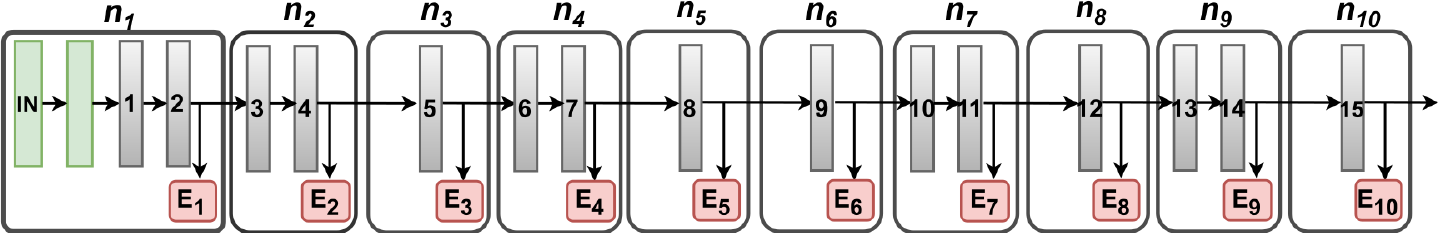}}
\end{center}
\caption{Distribution of exit points when layers and block of layers are distributed across different compute nodes. Red blocks represent the new exit points defined for the DNN models in \texttt{CONTINUER}.}
\label{fig:early-technique}
\vspace{-4.5mm}
\end{figure*}

\texttt{CONTINUER} adds exit points on ResNet-32 and MobileNetV2 under the assumption that each block of layers is placed across different nodes in the edge. Given that ResNet-32 can be distributed across 13 nodes, up to 13 different exit points can be added; one after each node ($n_{1}-n_{13}$) as shown in Figure~\ref{fig:ResNet-early}. The green bars represent individual layers, whereas the grey bars represent blocks distributed on thirteen nodes for ResNet-32. For MobileNetV2 there are 10 different positions where exit points can be added which are placed on up to 10 nodes ($n_{1}-n_{10}$) ( Figure~\ref{fig:MobileNet-early}).

In ResNet-32, an exit point is defined after each residual block. Each exit point comprises a convolutional layer with  $filter\:size=32, kernel\_size= 3, strides=2$, followed by a classifier that has a max pool layer, batch normalisation layer, and two dense layers of $units=64$, and $units=10$ respectively. The early exit points have the aforementioned layers so that prediction accuracy can be improved. The convolution layers specified at the exits points are fine-tuned based on experience 
to extract coarse level features of an input images that will be used by the classifers at the exit points. 

For MobileNetV2, the exit points are defined after the residual blocks represented as grey bars and numbered 2, 4, 5, 7, 8, 9, 11, 12, 14, and 15 in Figure~\ref{fig:early-technique}. The structure of the exit point defined for MobileNetV2 residual Block 2, includes a batch normalisation layer and then a convolutional layer with $filter\: size=96, kernel\_size= 3, strides=1$, followed by the classifier that has a global max pool layer and two dense layers of $units=64$ and $units=10$. A batch normalisation layer, followed by two convolutional layers with filter sizes of 160 and 80 are defined for residual blocks 4 and 5, followed by a classifier that has a global max pool layer and two dense layers of $units=64$ and $units=10$. For residual blocks 7, 8, 9, 11, and 12, a batch normalisation layer is defined followed by a convolutional layer with a filter size of 320, and a classifier that has a global maxing pool layer, and two dense layers of $units=64$ and $units=10$. For blocks 14, and 15, a batch normalisation layer, followed by a convolutional layer with $filter\: size=160, kernel\_size= 3,stride=1$ is defined, followed by a classifier layer that has a global max pool layer and two dense layers of $units=64$ and $units=10$. A convolutional layer with a stride of $size=1$ and a global max-pooling layer are employed based on experience and trial-and-error to improve the prediction performance of MobileNetV2. The batch normalisation layer is utilised to improve the training performance.

\subsubsection*{Model training parameters for Early-exit}
For training the DNN models that can make use of the early exit technique, a cross entropy loss function $L_{i}$ is employed for each of the early exit points $i = 1,..., N$ , and a total loss function $L_{T}$ is generated that is the weighted sum of these loss functions. 
The ResNet-32 model along with the intermediate exit points that are added is trained with a learning rate of $1e-3$ and for MobileNetV2, the model with exit points defined is with a learning rate of $1e-4$. Both models are trained with 500 epochs and a batch size of 64, on the CIFAR-10 dataset. 

\begin{figure}[t]
\begin{center}
	\subfloat[ResNet-32]
	{\label{fig:early-exit-accuracy-resnet}
	\includegraphics[width=0.237\textwidth]
	{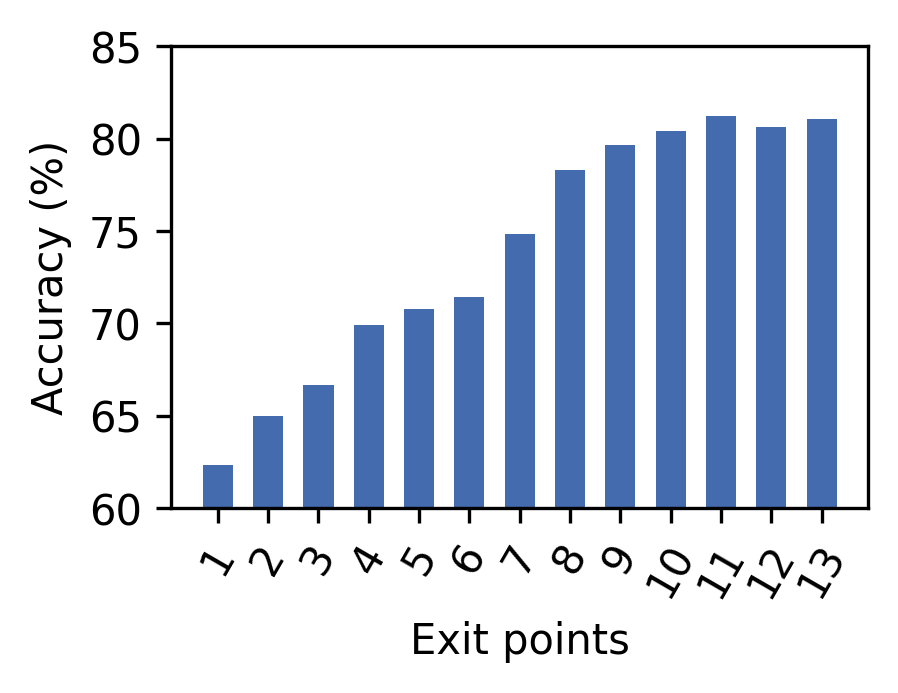}}
	\hfill
	\subfloat[MobileNetV2]
	{\label{fig:early-exit-accuracy-mobilenet}
	\includegraphics[width=0.237\textwidth]
	{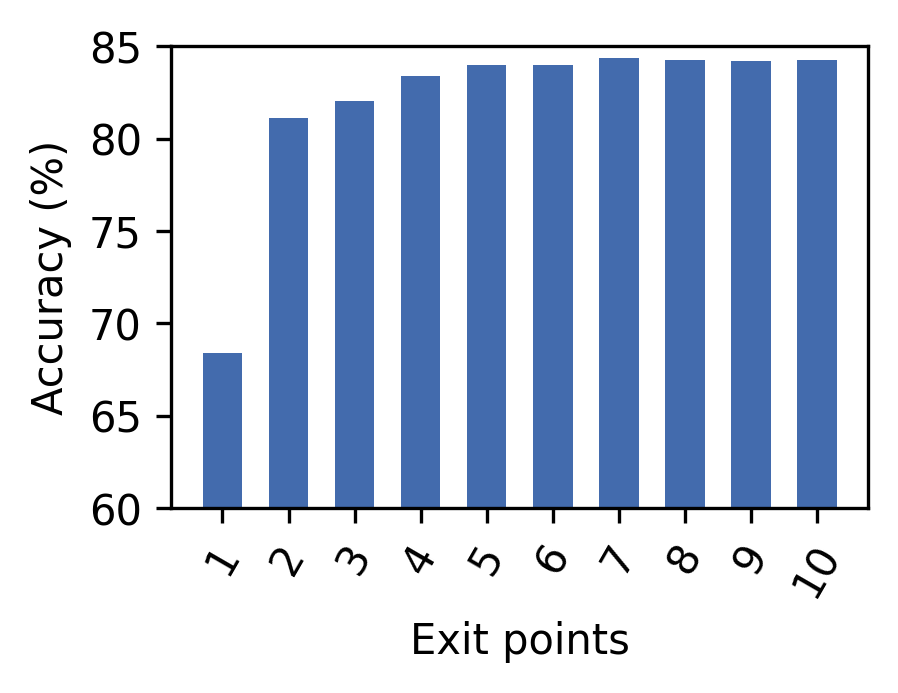}}
\end{center}
\caption{Accuracy of early exit points defined in ResNet-32 and MobileNetV2.}
\label{fig:early-exit-accuracy}
\vspace{-5mm}
\end{figure}

Figure~\ref{fig:early-exit-accuracy} shows the accuracy of ResNet-32 and MobileNetV2 for the different exit points. The x-axis shows the exit points and the y-axis shows the accuracy of the DNN models. As shown in the Figure~\ref{fig:early-exit-accuracy-resnet}, an accuracy lower than 70\% is noted for the initial early exit points ($E_1$ to $E_4$) ranging from 62.33\% to 69.92\%. Similarly, for MobileNetV2, $E_1$ has an accuracy of 68.39\%, this is to be expected. However, there is a trade-off between accuracy and the end-to-end latency of the DNN models when using different exit points, which will be presented in Section~\ref{sec:experiments}.

\subsubsection{Skip connection}
\begin{figure*}[t]
\begin{center}
	\subfloat[Skip connections defined for ResNet-32]
	{\label{fig:ResNet-skip}
	\includegraphics[width=0.75\textwidth]
	{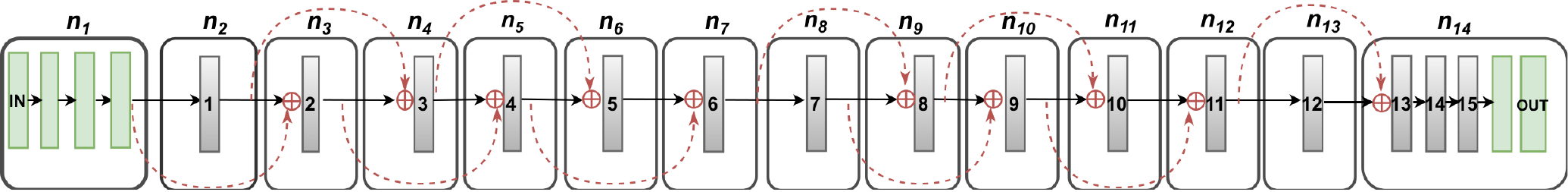}}
	
	\subfloat[Skip connections defined for MobileNetV2]
	{\label{fig:MobileNet-skip}
	\includegraphics[width=0.75\textwidth]
	{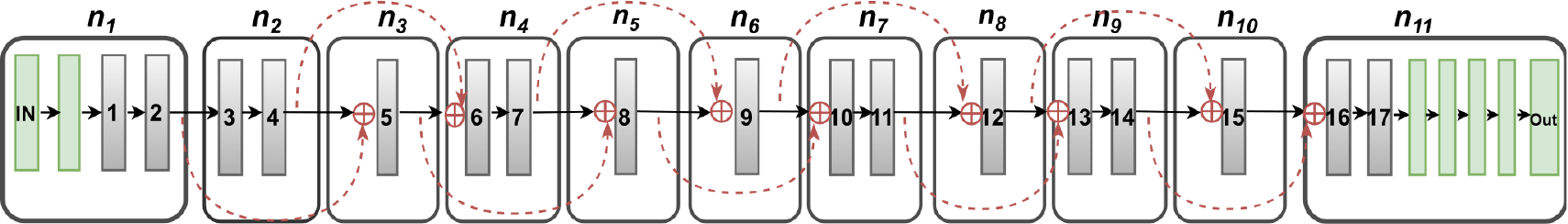}}
\end{center}
\caption{Skip connections for ResNet and MobileNetV2 models. The pink dotted lines are the skip connections added to the base model in \texttt{CONTINUER}.}
\label{fig:skip-technique}
\vspace{-4.5mm}
\end{figure*}

\texttt{CONTINUER} uses the default skip connections available in  ResNet-32 and MobileNetV2. Figure~\ref{fig:skip-technique} shows the skipping policy for skip connections for ResNet-32 and MobileNetV2. 
The grey bar represents a block of layers and the green bar represents a single layer. The red dotted line represents the skip connections added to the base model resulting in a total of 10 skip connections for ResNet-32 and 9 skip connections for MobileNetV2.

\subsubsection*{Model training parameters for Skip connection}
\begin{figure}[t]
\begin{center}
	\subfloat[ResNet-32]
	{\label{fig:skip-accuracy-resnet}
	\includegraphics[width=0.237\textwidth]
	{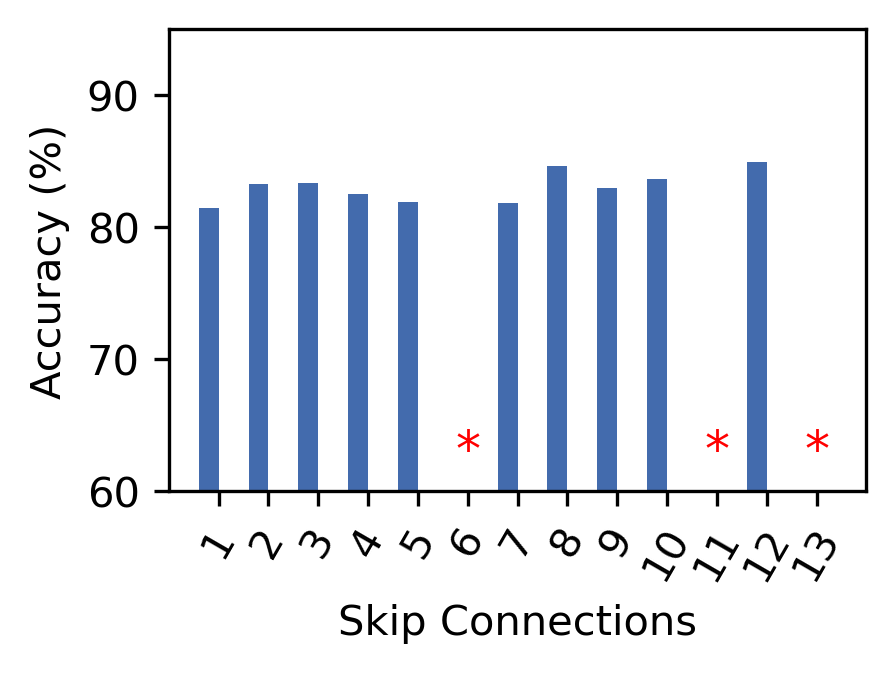}}
	\hfill
	\subfloat[MobileNetV2]
	{\label{fig:skip-accuracy-mobilenet}
	\includegraphics[width=0.237\textwidth]
	{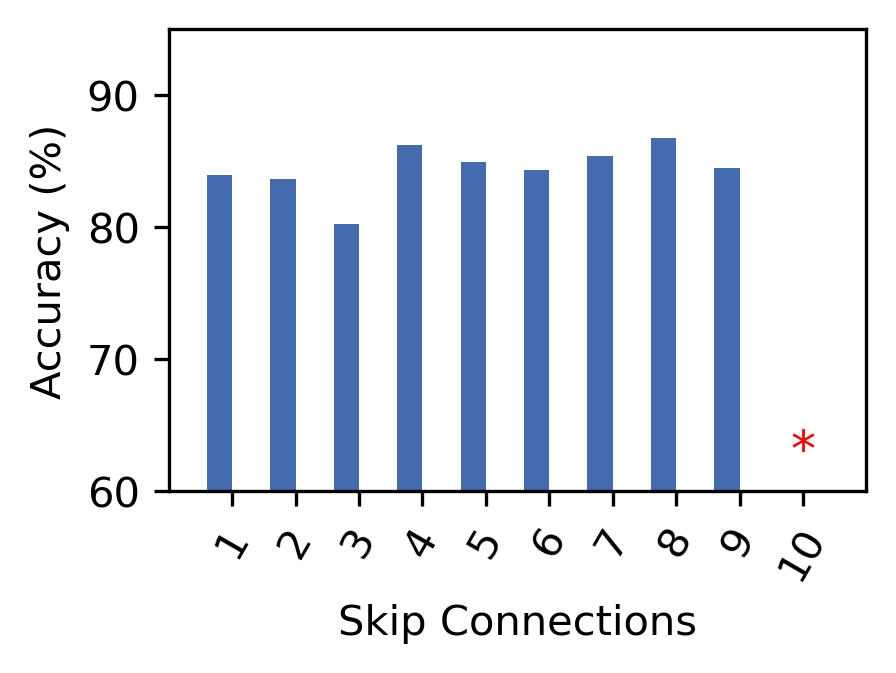}}
\end{center}
\caption{Accuracy of skip connections of ResNet and MobileNetV2 when using different skip connections. The red star indicates that skip connections are not possible at these points.}
\label{fig:skip-accuracy}
\vspace{-5mm}
\end{figure}

The learning rate is set to $1e-4$ for ResNet-32 and $1e-3$ for MobileNetV2. Both models are trained with a batch size of 64 and epoch size 500 using the categorical cross entropy as loss function. Residual blocks which have layers in the path of the skip connection are ignored and these blocks are represented by a  red star as shown in Figure~\ref{fig:skip-accuracy}. Figure~\ref{fig:skip-accuracy} shows the accuracy of ResNet-32 and MobileNetV2 for the skip connections defined at different positions. The x-axis shows the number of skip connections whereas the y-axis shows the accuracy. For ResNet-32, highest accuracy of 84.98\% is obtained for skip connection 12 whereas for MobileNetV2 an accuracy of 86.91\% is obtained for skip connection 8. The accuracy obtained for the skip connections defined in the ResNet-32 and MobileNetV2 indicates that skipping layers at runtime has a low impact on prediction accuracy. The settings for training hyperparameters defined for early-exit and skip connections are determined by trial and error.

The above three techniques can be extended to other DNN models. For DNN models that do not have default skip connections, skip connections need to be defined at specific locations in the DNN model and model retraining is required and similarly for early-exit technique.

\subsection{Metrics} 
The objective of the \texttt{CONTINUER} framework is to select the most suitable technique when a node fails while considering metrics such as end-to-end inference latency, accuracy and downtime. 
This section presents the methods adopted for obtaining the values of these metrics.

\subsubsection*{i) End-to-end Latency}
\texttt{CONTINUER} adopts a layer-wise approach that has been adopted in the literature to profile the inference latency of each layer of the DNN~\cite{wang2020perfnet,kang2017neurosurgeon}. Based on the values of profiled end-to-end latency, a Latency Prediction Model (as shown in Figure~\ref{fig:framework}) is developed for each type of layer by varying the layer's hyperparameters as shown in Table~\ref{tab:prediction-latency-hyperparam}. The Keras layers API\footnote{https://keras.io/api/layers/} is used for extracting the execution time of each layer type on two different processors that will be further discussed in Section~\ref{sec:experiments}. The advantage of using a layer-wise approach is that it profiles the inference latency of each type of layer instead of the whole DNN model, thus making it DNN-independent with a low profiling cost. 

\begin{table}[ht]
\centering
\caption{Layer hyperparameters of the DNN model.}
\label{tab:prediction-latency-hyperparam}
\begin{tabular}{|c|c|}
\hline
\textbf{\begin{tabular}[c]{@{}c@{}}Layer Type\end{tabular}} & \textbf{Hyperparameters}                                                                            \\ \hline
Batch Normalisation & input shape, input channel \\ \hline
Convolution                                                  & \begin{tabular}[c]{@{}c@{}}input shape, input channel, \\ kernel size, stride, filter\end{tabular} \\ \hline
ReLu                & input shape, input channel \\ \hline
Dense               & input shape, input channel \\ \hline
Add                 & input shape, input channel \\ \hline
Dropout             & input shape, input channel \\ \hline
Depthwise Convolution                                                 & \begin{tabular}[c]{@{}c@{}}input shape, input channel,\\  kernel size, stride\end{tabular}          \\ \hline
Global Pool Average & input shape, input channel \\ \hline
\end{tabular}
\vspace{-4mm}
\end{table}

To predict the end-to-end latency of the DNN, the Latency Prediction Model is trained using XGBoost~\cite{chen2016xgboost}. The hyperparameters of XGBoost are optimised using the Optuna\footnote{https://optuna.org/} optimisation library. The following best performing hyperparameters are identified by the optimisation framework for XGBoost $learning\:rate=0.1, n\_estimators=1000, max\:depth=10, colsample\, bytree=1, min child\: weight=1, seed=123$. 
The histogram-based algorithm is chosen as the XGBoost tree method. 
The quality of the predictions of the Latency Prediction Model is evaluated by quantifying the Mean Squared Error (MSE) and Coefficient of determination (denoted as $R^2$). Table~\ref{table:accuracy-layertype} shows the accuracy of each layer latency prediction models. The $R_{2}$ values except that of the dense layer, are close to 1 and a very low MSE is noted. This indicates that the Latency Prediction Model is fit for estimating layer latencies.

\begin{table}[t]
\centering
\caption{Accuracy of predicting the latency of different layer types}
\label{table:accuracy-layertype}
\begin{tabular}{|c|l|l|}
\hline
\textbf{\begin{tabular}[c]{@{}c@{}}Layer Type\end{tabular}} & \multicolumn{1}{c|}{\textbf{\begin{tabular}[c]{@{}c@{}} MSE \end{tabular}}} & \textbf{$R^2$} \\ \hline
Batch Normalistaion & 0.045 &	0.957 \\ \hline
Convolution &	0.040	& 0.980 \\ \hline
ReLu &	0.021	& 0.993 \\ \hline
Dense &	0.021	& 0.854 \\ \hline
Add & 0.009 &	0.995 \\ \hline
Dropout & 0.044 &	0.941 \\ \hline
Depthwise Convolution &0.008 &	0.994 \\ \hline
Global Pool Average &0.023 &0.992 \\ \hline
\end{tabular}
\vspace{-5mm}
\end{table}

\subsubsection*{ii) Accuracy}
Two approaches have been used in the literature to estimate the DNN model accuracy without profiling the input data.
The first  uses the model training parameters and hyperparameters~\cite{predictingDemogen}. The hyperparameters include the characteristics of the DNN architecture, such as  number of layers, and the training parameters include learning rate and loss function. The second approach involves predicting the accuracy of the DNN model by providing the pre-trained weights of the DNN model as input~\cite{2020predicting}. The training/test data or the meta-data of the DNN architecture may not be available in the production setting or when an edge node fails. Hence, estimating  DNN accuracy using the characteristics of the DNN is impractical. The second approach 
is thus adopted in \texttt{CONTINUER} by the Accuracy Prediction Model.

\begin{table}[ht]
\caption{Parameters for predicting accuracy}
\label{tab:parameters-training-acc}
\begin{tabular}{|c|c|}
\hline
\textbf{Parameters} & \textbf{Description}                               \\ \hline
Activation          & Activation function of the DNN model               \\ \hline
B\_init             & Initialisation used for bias weights               \\ \hline
DNN\_architecture   & DNN model (ResNet32, MobileNetV2)       \\ \hline
Epochs              & Total number of epochs (500)                       \\ \hline
Learning\_rate      & Learning rate                                      \\ \hline
num\_layers         & Number of layers                               \\ \hline
Optimiser           & Optimiser used (Adam optimiser)                    \\ \hline
Train\_fraction     & Fraction of the total number of training samples \\ \hline
Train\_accuracy     & Accuracy on the training set                       \\ \hline
Train\_loss         & Loss on the training                               \\ \hline
\end{tabular}
\vspace{-4mm}
\end{table}

Table~\ref{tab:parameters-training-acc} shows the parameters used when training the Accuracy Prediction Model for ResNet-32 and MobileNetV2.
The parameters are extracted by implementing a custom Keras callback function that is invoked at the end of each epoch during training. 

The Accuracy Prediction Model is based on the Light Gradient Boosting Machine (LightGBM)~\cite{ke2017lightgbm} that predicts accuracy by providing the pretrained weights of the DNN model. The settings defined for LightGBM hyperparameters are the following: $learning\:rate=0.1, n\_estimators=100, max\:depth=-1, colsample\, bytree=1.0, min child\: weight=0.001$.
A split ratio of 80:20 is used to split the data into training and testing data. The pre-trained weights provided to LightGBM are pre-processed by using the mean, variance, and $q^{th}$ percentiles for $q \epsilon \left \{ 0, 25, 50, 75, 100\right \}$ for each layer of the DNN model as presented in the literature~\cite {2020predicting}. The ResNet-32 model is trained with a learning rate set to $1e-3$ for all the three techniques. For the repartitioning and skip connection techniques, the learning rate is set to $1e−3$ for MobileNetV2 and to $1e-4$ in early-exit. Both models are trained using the loss function categorical cross-entropy on the CIFAR-10 dataset with a batch size of 64 and epoch size of 500. Model training parameters are determined by trial and error.
The MSE obtained is 0.223 (low indicates that the estimation has a high accuracy) and $R^2$ is 98.01\% (high indicates that there is a high correlation between the input parameters to the prediction model and the output).

\subsubsection*{iii) Downtime}
The downtime is the time taken to retrieve the estimated accuracy and latency from the Accuracy Prediction Model and Latency Prediction Model, respectively and for the Scheduler (discussed further in Section~\ref{subsec:runtimephase} to select one of the three techniques, namely repartitioning, early-exit and skip connection when an edge node fails. 
The repartitioning and skip connection techniques have an additional 0.99ms downtime to reinstate connections~\cite{majeed2021neukonfig}. 

\subsection{Runtime phase} 
\label{subsec:runtimephase}

The Scheduler is a key component used in the runtime phase. The Scheduler takes as input the estimated accuracy and the estimated latency of the DNN for all three techniques and the downtime (empirically obtained) that will be incurred by each technique and determines the suitable technique for node failure. 
The accuracy is obtained from the Accuracy Prediction Model, the end-to-end latency is obtained from the Latency Prediction Model. 
The goal of the Scheduler is to select the technique that best satisfies any user provided objectives, such as thresholds for end-to-end latency, accuracy and downtime. 

To minimise the cost of selecting a suitable technique subject to user-defined objectives, a classic additive weighting method is employed~\cite{yoon1995multiple}. Different weights are assigned to each objective and then the weighted sum of the normalised values of the user-defined objectives is minimised. If the accuracy, end-to-end latency and downtime objectives are denoted as $A$, $L$ and $D$, respectively, then the normalised objectives are denoted as $A'$, $L'$ and $D'$ (normalised to a value between 0 and 1 using the Linear Max-Min  technique).

The suitable selection of technique considering user requirement is formulated in Equation~\ref{eqn:optimisation}.
\begin{equation}
min \sum\omega_{1}A'-\omega_{2}L'-\omega_{3}D'
\label{eqn:optimisation}
\end{equation}
where $\omega_{1}, \omega_{2}, \omega_{3} $ are the dynamic weights of accuracy, end-to-end latency, and downtime. The value of each weight factor, represents the level of importance, and is set by the user. For instance, if the user has not specified a latency threshold, a weight factor of 0 is assigned to latency objective.

\section{Experimental Studies}
\label{sec:experiments}
This section  presents the experimental setup used to assess the \texttt{CONTINUER} framework and the results obtained. 

\subsection{Experimental Setup}
Experimental studies are carried out on two 64-bit x86 processor platforms as shown in Table~\ref{tab:platform}; similar processors have been used as representative of edge environments in the literature~\cite{abouaomar2021resource,puliafito2019fog}. Since latency is resource dependent, the results of the Latency Prediction Model are obtained from both platforms. Accuracy is not impacted by the platforms. 
The repartitioning, early-exit, and skip connection techniques are examined in the context of two DNNs in \texttt{CONTINUER}, namely ResNet-32, and MobileNetV2. The DNNs are implemented using the Tensorflow library and the Keras API.  

\begin{table}[t]
\centering
\caption{Processor platforms used in the experimental studies.}
\label{tab:platform}
\begin{tabular}{|l|l|l|l|}
\hline
\multicolumn{1}{|c|}{\textbf{Platform}} & \multicolumn{1}{c|}{\textbf{CPU}} & \multicolumn{1}{c|}{\textbf{Clock Freq.}} & \multicolumn{1}{c|}{\textbf{Memory}} \\ \hline
\textbf{Platform 1} & Intel(R) Core (TM) i7-8700& 3.20GHz & 16GB\\ \hline
\textbf{Platform 2}& Intel(R) Core (TM) i5-8250U& 1.60GHz & 16GB \\ \hline
\end{tabular}
\vspace{-5mm}
\end{table}

\subsection{Results}
The performance of the three key components in the \texttt{CONTINUER} framework, namely the Latency Prediction Model, the Accuracy Prediction Model and the Scheduler will be evaluated in this section. It will be demonstrated that the latency of the DNNs (resource dependent) can be estimated on both platforms with a high accuracy. Similarly, accuracy (resource independent) of the DNN model can be estimated with a high accuracy without the need for profiling the DNNs when a node fails. The results will also highlight that the Scheduler that is based on estimated data will select a suitable technique given user-defined objectives with a high accuracy.

\subsubsection{Quality of estimating latency and accuracy metrics}
The quality of the results estimated by the Latency Prediction Model and Accuracy Prediction Model is evaluated by comparing the estimated latency and accuracy with the measured latency and accuracy. Measured accuracy is obtained from the trained ResNet-32 and MobileNetV2 on the CIFAR-10 dataset for the three techniques. Measured latency is obtained by profiling the trained ResNet-32 and MobileNetV2 DNNs for the three techniques.

\begin{figure*}[t]
\begin{center}
	\subfloat[ResNet-32 on Platform 1]
	{\label{fig:resnet-platform1}
	\includegraphics[width=0.25\textwidth]
	{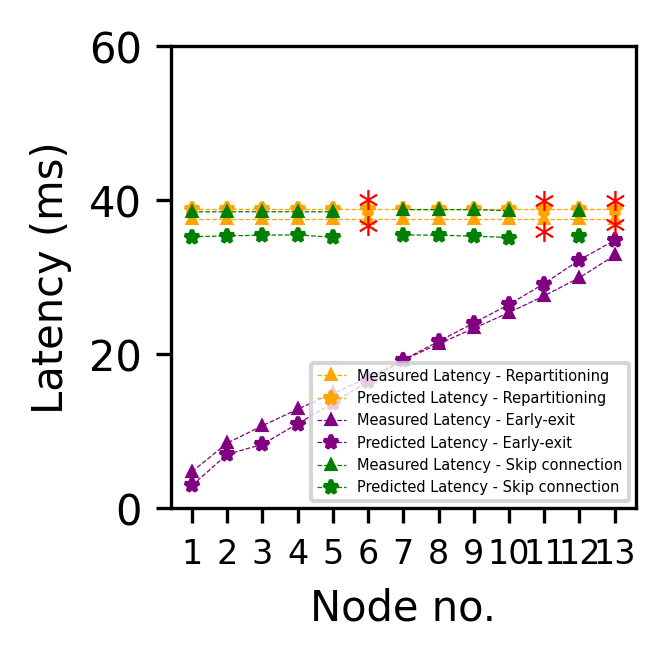}}
	\subfloat[ResNet-32 on Platform 2]
	{\label{fig:resnet-platform2}
	\includegraphics[width=0.25\textwidth]
	{sections/images/updated/lat-pred-resnet-plat1.png}}
	\subfloat[MobileNet-V2 on Platform 1]
	{\label{fig:mobilenet-platform1}
	\includegraphics[width=0.25\textwidth]
	{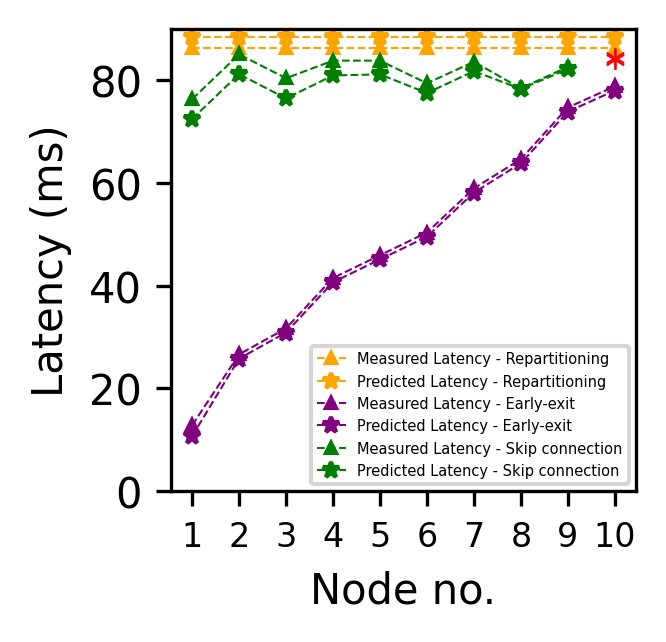}}
	\subfloat[MobileNet-V2 on Platform 2]
	{\label{fig:mobilenet-platform2}
	\includegraphics[width=0.25\textwidth]
	{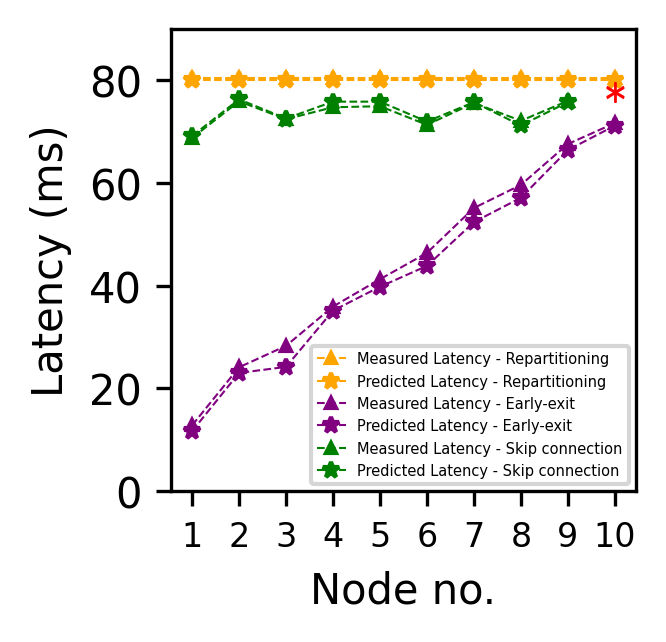}}
\end{center}

\caption{Measured and predicted latency for ResNet-32 and MobileNetV2.}
\label{fig:node-wise-latency-measured-pred}
\vspace{-5mm}
\end{figure*}

Figure~\ref{fig:node-wise-latency-measured-pred} shows the measured and predicted latency on different platforms for ResNet-32 and MobileNetV2. 
The x-axis shows the node number on which a block of DNN model is deployed and the y-axis shows the latency  of each of the techniques (repartitioning, early-exit, and skip connection). For repartitioning the latency is a constant for all nodes since the entire DNN is repartitioned and redeployed once an edge node fails. For early-exit, the inference request will be completed before a failed node through an exit point. Hence, latency for early-exit is the execution time of the request, which increases when the inference request passes through a larger number of nodes. For skip connection, the latency metric is the end-to-end latency of entire DNN deployed on edge nodes excluding the failed node (the node that was bypassed by the skip connection). 

For the skip connection, red stars indicate nodes on which a skip connection is not possible. Figure~\ref{fig:node-wise-latency-measured-pred} shows the measured and predicted latency of each technique. Table~\ref{tab:percentage-error-latency} shows the average percentage error of each technique for ResNet-32 and MobileNetV2. The percentage error is the difference between the measured and actual values of latency and accuracy of the three techniques. 
It is noted that the average percentage error for the three methods is relatively low; the maximum is 13.06\% 
for the early-exit technique on ResNet-32. 

\begin{figure*}[h!]
\begin{center}
	\subfloat[ResNet-32]
	{\label{fig:resnet-acc-meas-pred}
	\includegraphics[width=0.25\textwidth]
	{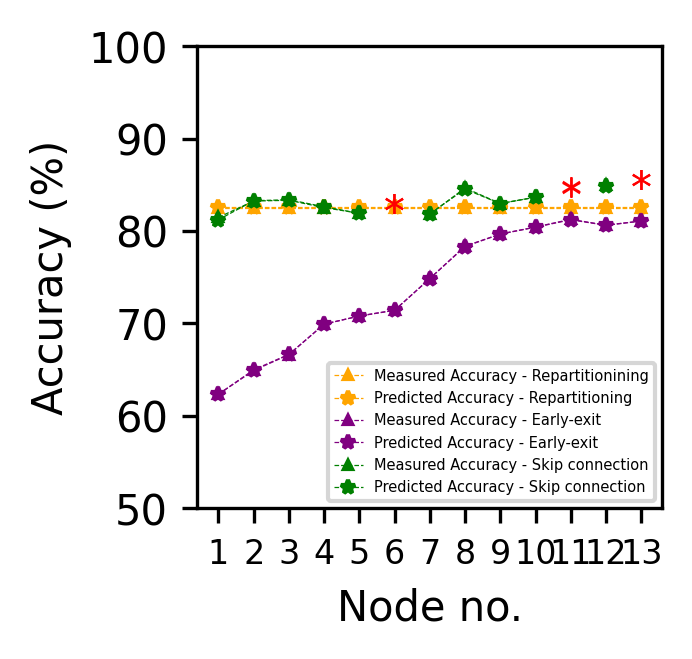}}
	\subfloat[MobileNetV2]
	{\label{fig:mobilenet-acc-meas-pred}
	\includegraphics[width=0.25\textwidth]
	{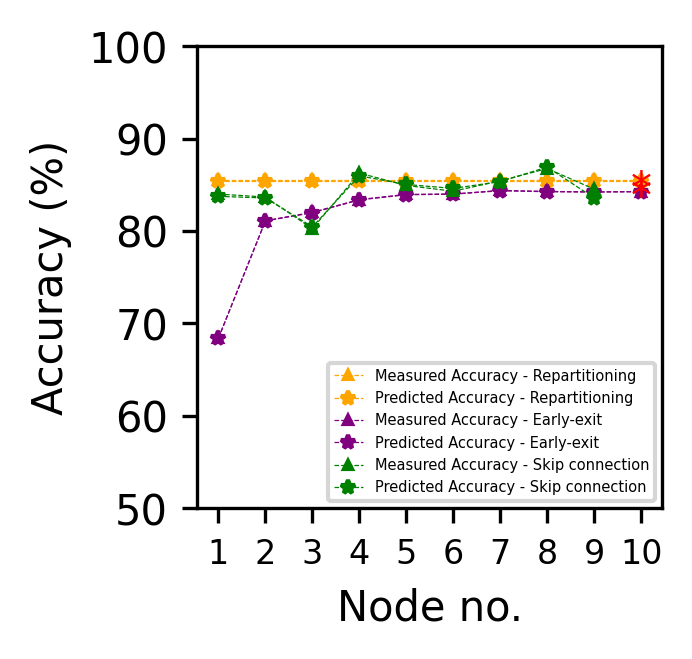}}
\end{center}
\caption{Measured and predicted accuracy for ResNet-32 and MobileNetV2.}
\label{fig:node-wise-acc-measured-pred}
\vspace{-5mm}
\end{figure*}

\begin{table}[t]
\centering
\caption{Average percentage error when estimating latency using the repartitiong, early-exit and skip connection techniques for ResNet-32 and MobileNetV2.}
\label{tab:percentage-error-latency}
\resizebox{8.7cm}{!}{%
\begin{tabular}{|l|cc|cc|}
\hline
\multirow{2}{*}{\textbf{Technique}} & \multicolumn{2}{c|}{\textbf{Platform 1}} & \multicolumn{2}{c|}{\textbf{Platform 2}} \\ \cline{2-5} 
 & \multicolumn{1}{l|}{\textbf{ResNet-32}} & \multicolumn{1}{l|}{\textbf{MobileNetV2}} & \multicolumn{1}{l|}{\textbf{ResNet-32}} & \multicolumn{1}{l|}{\textbf{MobileNetV2}} \\ \hline
\textbf{Repartitioning} & \multicolumn{1}{c|}{3.48\%} & 2.44\% & \multicolumn{1}{c|}{2.33\%} & 0.51\% \\ \hline
\textbf{Early-exit} & \multicolumn{1}{c|}{10.10\%} & 3.22\% & \multicolumn{1}{c|}{13.06\%} & 5.07\% \\ \hline
\textbf{Skip connection} & \multicolumn{1}{c|}{2.92\%} & 2.92\% & \multicolumn{1}{c|}{3.06\%} & 0.73\% \\ \hline
\end{tabular}%
}
\vspace{-5mm}
\end{table}

Figure~\ref{fig:node-wise-acc-measured-pred} shows the measured and predicted accuracy of ResNet-32 and MobileNetV2. The x-axis shows the node number on which a block of DNN model is deployed, whereas the y-axis shows the measured and estimated percentage accuracy of the DNN. The accuracy of ResNet-32 and MobileNetV2 remains the same on each node for repartitioning. For early-exit the accuracy of ResNet-32 increases for higher node number, whereas for MobileNetV2 the accuracy is 68.39\% for the first exit point and a higher accuracy is noted at subsequent exit points. The accuracy of ResNet-32 and MobileNetV2 varies slightly for the skip connection.

Table~\ref{tab:percentage-error-accuracy} shows the average percentage error of accuracy metric for ResNet-32 and MobileNet-V2 for each technique. The accuracy metric is more precisely estimated than the latency metric. The maximum average percentage error of 0.28\% is noted for skip connection.

\begin{table}[t]
\centering
\caption{Average percentage error when estimating accuracy using the repartitiong, early-exit and skip connection techniques for ResNet-32 and MobileNetV2.}
\label{tab:percentage-error-accuracy}
\begin{tabular}{|l|c|c|}
\hline
\textbf{Technique}       & \textbf{ResNet-32} & \textbf{MobileNetV2} \\ \hline
\textbf{Repartitioning}  & 0\%                  & 0.12\%               \\ \hline
\textbf{Early-exit}      & 0.03\%             & 0.03\%               \\ \hline
\textbf{Skip connection} & 0.06\%             & 0.28\%               \\ \hline
\end{tabular}
\vspace{-3.5mm}
\end{table}

\subsubsection{Quality of selection by the Scheduler}
To determine the quality of the technique selected by the Scheduler a parameter sweeping analysis is performed. There are three parameters to sweep, these are the weights for the accuracy, latency, and downtime ${\omega}={\omega_A,\omega_L,\omega_D}$. The constraint for all three parameters is defined as $0.1\geq \omega \leq 0.9 $ (in increments of 0.1).
Different combinations of weights are applied on each instance (dataset generated using the normalised values for estimated accuracy, estimated latency and downtime for all techniques and for all nodes) for ResNet-32 and MobileNetV2. Min–Max normalisation is applied on the accuracy, latency and downtime metric to ensure the accuracy of the selection of the Scheduler. Then the objective function defined in Equation~\ref{eqn:optimisation} takes as input the estimated accuracy, estimated latency and downtime of each technique to determine which technique is suitable when an edge node fails. Parameter sweeping is applied on the normalised dataset which results in 767 instances for ResNet-32 and 590 instances for MobileNetV2. 
The quality of the selection made is based on the classification accuracy metric, which is the total number of correctly identified techniques divided by the total number of instances. 

\begin{table}[t]
\centering
\caption{Quality of the selection made by the \texttt{CONTINUER} Scheduler when selecting a suitable technique.}
\label{tab:scheduler-result}
\begin{tabular}{|l|l|l|}
\hline
\multicolumn{1}{|c|}{\textbf{DNN Model}} & \multicolumn{1}{c|}{\textbf{Platform 1}} & \multicolumn{1}{c|}{\textbf{Platform 2}} \\ \hline
\textbf{ResNet-32}                       & 99.86\%                                  & 99.86\%                                  \\ \hline
\textbf{MobileNetV2}                     & 86.12\%                                  & 99.83\%                                  \\ \hline
\end{tabular}
\vspace{-5mm}
\end{table}

Table~\ref{tab:scheduler-result} shows that the appropriate technique is selected by the Scheduler with up to an accuracy of 99.86\%. The results demonstrate that the Scheduler is effective in determining the appropriate technique by considering user-defined objectives.

\subsubsection{Overhead}
Table~\ref{tab:downtime-result} shows the overhead (maximum downtime) incurred for the repartitioning, early-exit and skip connection in milliseconds (ms). Downtime is the sum of the time taken to retrieve the estimated accuracy and end-to-end latency metrics and to select a suitable technique based on the user-defined objectives. \texttt{CONTINUER} selects a suitable technique within 16.82~ms following a node failure.

\begin{table}[t]
\centering
\caption{Downtime incurred when selecting a technique.}
\label{tab:downtime-result}
\begin{tabular}{|l|c|c|}
\hline
\multicolumn{1}{|c|}{\textbf{Technique}} & \multicolumn{1}{l|}{\textbf{ResNet-32}} & \multicolumn{1}{l|}{\textbf{MobileNetV2}} \\ \hline
\textbf{Repartitioning}  & 3.56ms & 16.16ms \\ \hline
\textbf{Early-exit}      & 1.83ms & 9.28ms  \\ \hline
\textbf{Skip connection} & 3.32ms & 16.82ms \\ \hline
\end{tabular}
\vspace{-5mm}
\end{table}

\subsubsection*{Summary}
The above results highlight that the Accuracy Prediction Model and Latency Prediction Model of \texttt{CONTINUER} estimate the accuracy and latency metric with a relatively low average percentage error. An accuracy of up to 99.86\% is obtained in selecting a suitable technique when an edge failure occurs with a relatively low overhead. These confirm the feasibility of \texttt{CONTINUER}.

\section{Related Work}
\label{sec:relatedwork}

Two aspects relevant to the research reported in this paper are considered in this section. The first is approaches that address runtime concerns when distributed DNNs are deployed; for example, adapting to variable network speeds and resource availability of compute resources. This is relevant as maintaining services when an edge failure occurs is a runtime concern. 
The second is consideration of edge failures. 

Repartitioning is used to address runtime concerns, such as change in network speed and resource variability, by repartitioning and redeploying DNNs to suit the given operational conditions in NEUKONFIG~\cite{majeed2021neukonfig}. Approaches are incorporated to reduce the service downtime when repartitioning.

Early-exit has been used in the context of addressing the runtime concern of varying network speeds and availability for distributed DNNs (e.g, Edgent~\cite{li2019edgent}). Similarly, the latency is reduced in the context of industrial IoT environments using the early-exit technique in Boomerang~\cite{zeng2019boomerang}. 

Skip connection is used for reducing the inference latency by introducing a layer-wise skipping policy in SkipNet~\cite{skipnet}. Supervised and reinforcement learning approaches are used to improve the policy by conditioning it against the input sample. Similar approaches such as BlockDrop~\cite{wu2018blockdrop} skip residual blocks at runtime to reduce inference delay.

Existing research that addresses edge failures of DNN services has considered the early-exit approach. One such example is SEE~\cite{wang2019see} in which it is assumed that the duration of the failure is known apriori. This is impractical for a dynamically changing environment such as the edge. LEE fills this gap of SEE by knowing the duration of failure~\cite{ju2021learning}. However, SEE and LEE make the decision for individual video frames and uses buffers to hold pending inference requests. The decision made for a current frame may negatively impact future frames.
Another example is DeepFogGuard~\cite{yousefpour2019guardians} in which skip connection is used to skip failed physical nodes on which a distributed DNN is deployed. However, the approach used requires retraining the DNN which makes it less suitable for responding to runtime changes, such as the failure of an edge node. In addition, the work does not quantify the downtime incurred by skipping connections during failure. 
Generally, the above research does not account for the resource limited nature of the edge environment in response to a service outage. Further, existing research does not consider the rapid response to an edge failure that may be required for latency-critical applications relying on DNNs. 
They do not quantify the overhead that will be incurred and consider user-defined objectives when an edge node fails.

\section{Conclusions}
\label{sec:conclusions}
This paper presents the \texttt{CONTINUER} framework to maintain the service of distributed DNNs when an edge outage occurs. \texttt{CONTINUER} is underpinned by three techniques, namely repartitioning, early-exit and skip connection, one of which is selected by the framework to continue delivering DNN services when an edge node fails by accounting for trade-offs in accuracy, end-to-end latency and downtime and user-defined objectives. Experiments on a lab-based testbed show that \texttt{CONTINUER} selects the appropriate technique with high accuracy and low overhead, thus confirming the feasibility of the framework. 

\textit{Limitations and Future Work:}
Despite the challenge addressed by \texttt{CONTINUER}, it has certain limitations. Firstly, for estimating the end-to-end latency of the DNN, the approach for profiling DNN layers is CPU oriented. Future work will include GPU-aware estimation of latency.
Secondly, \texttt{CONTINUER} considers ResNet-32 and MobileNetV2, which by default have skip connections defined in them. The research can be expanded to consider other DNNs. Thirdly, it is assumed in this paper that each DNN block is deployed on a different edge node and only a single node fails. Efforts will be made to address these limitation in the future. 

\bibliographystyle{IEEEtran}  
\bibliography{paper-v1}

\end{document}